\documentclass{article}
\usepackage{graphicx} % Required for inserting images
\usepackage{amsmath}
\usepackage{xcolor}
\usepackage[justification=centering]{caption}

\usepackage[english]{babel}
\usepackage[titles]{tocloft}

\usepackage{amsmath}
\usepackage{graphicx}
\usepackage{algorithm}
\usepackage{algpseudocode}
\usepackage{enumitem}
\usepackage{amsmath}
\usepackage{amssymb}
\usepackage{booktabs}
\usepackage{frontespizio}
\usepackage{listings}
\usepackage[square,numbers]{natbib}
\usepackage[export]{adjustbox}

\usepackage{hyperref}
\hypersetup{
    colorlinks=true,
    allcolors=blue,
    pdftitle={Integrating Multiple Data Sources with Interactions in Multi-Omics Using Cooperative Learning},
    pdfpagemode=FullScreen,
}

\title{Integrating Multiple Data Sources with Interactions in Multi-Omics Using Cooperative Learning}
\author{Matteo D'Alessandro$^{*}$, Theophilus Quachie Asenso, Manuela Zucknick}
\date{Oslo Centre for Biostatistics and Epidemiology, Department of Biostatistics, University of Oslo\\%
    $^*$\textit{matteo.dalessandro@medisin.uio.no}\\[2ex]%
    \today}

\begin{document}

\maketitle

\begin{abstract}
Modeling with multi-omics data presents multiple challenges such as the high-dimensionality of the problem ($p \gg n$), the presence of interactions between features, and the need for integration between multiple data sources. We establish an interaction model that allows for the inclusion of multiple sources of data from the integration of two existing methods, pliable lasso and cooperative learning. The integrated model is tested both on simulation studies and on real multi-omics datasets for predicting labor onset and cancer treatment response. The results show that the model is effective in modeling multi-source data in various scenarios where interactions are present, both in terms of prediction performance and selection of relevant variables.\\

\textit{Keywords}: Cooperative learning, pliable lasso, interaction models, multi-omics, personalized medicine.
\end{abstract}

\section{Introduction}

In recent years, high-dimensional omics data have become increasingly prevalent in biomedical research, with the advent of data-driven disciplines such as genomics, transcriptomics, proteomics, and metabolomics. These technologies allow for the simultaneous measurement of thousands or even millions of molecular features, generating vast amounts of data that provide insights into the underlying biological processes and mechanisms of disease.

Multiomics analysis offers a powerful approach to gain a holistic understanding of biological systems. By integrating data from multiple molecular layers, it allows for effective investigation of biomarkers, pathways, and mechanisms behind various biological processes. It also plays a crucial role in personalized medicine by assessing health status, predicting disease susceptibility, and guiding targeted treatments. Multiomics studies have achieved remarkable results in tumor classification, prognostic predictions in cancer research, and the discovery of drug targets for various diseases \citep{Lindskrog2021, Joshi2020}.\\

However, the analysis of this form of data presents significant statistical and computational challenges. The large number of variables involved, as well as the high degree of correlation among them introduce problems that are not present in more low-dimensional settings. In this scenario, traditional statistical methods may not be applicable or may lead to unreliable results. Penalized regression methods such as lasso \citep{Tibshirani1996} or elastic net \citep{Zou2005} are often used in this setting to perform both variable selection and estimation of the coefficients simultaneously, and with computationally efficient algorithms.
Moreover, failing to account for interactions between variables can result in a limited understanding of the complex relationships within the data, as interactions can uncover synergistic or antagonistic effects that may not be captured by main effects alone. An issue when considering interactions is the high number of parameters this introduces in the model: if we aim at considering all possible interaction pairs, the interaction effects would grow quadratically with the number of main effects. A common approach to tackling this issue is by introducing interactions in a hierarchical manner: this way, for an interaction term to be entered, it is required for either one or both associated main effects to also be included in the model. This hierarchy can be imposed through setting convex constraints on the solution \citep{Bien2013}, or more commonly through a variation on the group-lasso penalty (\cite{Zhao2009,Lim2015,Choi2010}, among others). A member of this latter class of models, the pliable lasso \citep{Tibshirani2019}, considers interactions between main predictors and a set of modifying variables, only allowing an interaction term to be entered if the corresponding main-effect is non-zero.\\

A second challenge related to the use of omics data is the need for integration between different data sources. The combination of multiple omics platforms, such as gene expression, mutations, copy number variations or protein abundance, can help researchers identify patterns and relationships that may not be apparent from a single source of data. The integration can provide a more complete picture of biological processes and help identify personalized treatment options.
Multiple statistical methods for multi-source problems have been proposed so far. Two straightforward approaches are early and late fusion. Early (or feature-level) fusion begins by transforming all datasets into a single representation, which is then used as the input to a supervised learning model of choice. Late (or prediction-level) fusion, instead, works by developing first-level models from individual data views and then combining the predictions by training a second-level model as the final predictor \citep{Boulahia2021}. A 
different approach is taken in \cite{Zhao2014}, which involves preselecting a small number of variables from each source through Lasso, to use in a second modeling step, while the IPF-Lasso \citep{Boulesteix2017} represents an extension of Lasso to account for multiple sources that require different amounts of penalization. Another class of methods relies on the introduction of an ``alignment term" in the cost function, that encourages the prediction from different data views to be similar: methods in this group include collaborative regression \citep{Gross2014} and cooperative learning \citep{Ding2022}.\\

Our aim is to develop an extension of a hierarchical interaction model, the pliable lasso, to allow it to consider multiple data sources in the context of cooperative learning. The dual ability of this model to deal with information coming from different sources and to differentiate the coefficients on the basis of specific variables would make it especially fit when working with biological data.

The remainder of the article is structured as follows. In Section \ref{section_method} the relevant methods are introduced, and the details of our proposal are presented. The model is then tested on simulated data (Section \ref{simulation_studies_section}) both in a high-dimensional and low-dimensional setting, and on real multi-omics datasets (Section \ref{section_real_multi-omics_studies}) concerning labor onset and cancer treatment response prediction. A final discussion of the results and some thoughts for future research are included in Section \ref{section_conclusion}.\\

\section{Methods}\label{section_method}

\subsection{Pliable Lasso}

The pliable lasso model, first introduced by \cite{Tibshirani2019}, is a generalization of the usual lasso allowing the model coefficients to vary as a function of a set of modifying variables. An example of modifying variable in a biological setting could be sex, meaning that some main effect coefficients could differ between male and female subjects.

Letting $y$ be the $n$-vector of outcomes, $X$, $Z$ be $n \times p$ and $n \times K$ matrices of predictors and modifying variables respectively, $X_j$ be the $j$-th column of $X$, and \textbf{1} be a column $n$-vector of ones, the estimation has the form:

\begin{equation} \label{eq_pliable}
\begin{split}
\hat{y} & =  \hat{\beta_0} \textbf{1} + Z\hat{\theta_0} + \sum_{j=1}^p X_j(\hat{\beta_j} \textbf{1} + Z\hat{\theta_j}) \\
 & =  \hat{\beta_0} \textbf{1} + Z\hat{\theta_0} + \sum_{j=1}^p (X_j\hat{\beta_j} + W_j\hat{\theta_j})\end{split},
\end{equation} 
where $W_j = X_j \circ Z$ with $(\circ)$ representing elementwise multiplication.
An asymmetric weak hierarchy constraint between the coefficients is added: $\theta_j$ can be non-zero only if $\beta_j$ is non-zero. 
In order to enforce this constraint and give sparsity to the components of $\theta_j$, the objective function considered for the problem is:

\begin{equation} \label{cost_function_pliable}
\begin{split}
J(\beta_0,\theta_0,\beta,\theta)& = \frac{1}{2N}\sum_{i=1}^n(y_i-\hat{y_i})^2\\
& + (1-\alpha)\lambda\sum_{j=1}^p(||(\beta_j,\theta_j)||_2 + ||\theta_j||_2)\\
& + \alpha\lambda\sum_{j,k}|\theta_{jk}|,\end{split}
\end{equation}
where $\alpha$ and $\lambda$ are tuning parameters. The penalty terms are similar to those introduced in the sparse group lasso.
The first term enforces the hierarchy constraint; the non-differentiable point in the non-squared $\ell_2$-norm gives sparsity both to the coupled coefficients $(\beta_j,\theta_j)$ and to the single vector $\theta_j$: this way, if $\beta_j$ is zero, $\theta_j$ is also forced to be zero.
The second term introduces sparsity in the components of $\theta_j$ as for the usual lasso penalty. The main advantage in the use of pliable lasso therefore lays in the hierarchical inclusion of interaction variables, which first considers whether the corresponding main effect has been included, which helps the model avoid over-fitting.

As highlighted in detail in the original paper \citep{Tibshirani2019}, the objective function (\ref{cost_function_pliable}) for the problem is convex and can be minimized through a blockwise cyclical coordinate descent procedure. Explicit conditions are introduced to determine whether $(\hat{\beta}_j,\hat{\theta}_j)$ is nonzero and if it is, if the $\hat{\theta_j}$ component is nonzero. Only in this last case, both parameters are determined through gradient descent. This allows for a faster computation, since the costlier gradient descent is only performed when needed. The screening conditions can be obtained through the sub-gradient equations of (\ref{cost_function_pliable}) (see \cite{Tibshirani2019} for details of the optimization).\\

After its proposal, the modeling of interactions through the pliable lasso penalty has been expanded to other methods. In particular, some applications include the Cox proportional hazards model framework \citep{PliableCox}, the classification setting with multinomial logistic regression \citep{Asenso2020}, the support vector machine \citep{Asenso2022} and the multi-response regression problem \citep{https://doi.org/10.48550/arxiv.2303.11155}.
Some real data examples that the pliable lasso has been applied to include HIV mutation data, skin cancer proteomics data and stock returns prediction \citep{Tibshirani2019}.

\subsection{Cooperative learning}

We consider the context of supervised learning with $m$ datasets $X_1, \dots X_m \in \mathbb{R}^{n\times p}$, containing information from multiples data sources. The goal of data fusion techniques is to integrate these sources to make a prediction on a response variable $y$.
Data fusion plays a crucial role in the field of sensors and imaging, as it enables the integration of information from different sensors or imaging modalities, such as radar and optical sensors, which can improve the overall quality and reliability of the data \citep{Zhang2010}.
In medical imaging, data fusion can be used to combine data from different imaging modalities, such as MRI and CT scans, to obtain a more accurate diagnosis \citep{Adali2015}. In the context of multiomics data, this type of integration can provide insights into complex biological processes that cannot be elucidated by any single omics platform alone \citep{Leng2022}.\\

The cooperative learning method \citep{Ding2022}, in the case of two data views, aims to solve the following minimization problem:

\begin{equation} \label{min_prob_cooplearn}
    \text{min    } E[ \frac{1}{2} (\mathbf{y} - f_1(X_1) - f_2(X_2))^2 + \frac{\rho}{2}(f_1(X_1) - f_2(X_2))^2 ].
\end{equation}

The cost function combines the usual squared error loss of predictions with an “agreement” penalty that encourages the predictions to align. The approach proves to be especially useful when the data sources share some underlying information that, when considered jointly, can allow for better prediction by strengthening the aligned signal across modalities. By varying the weight of the agreement penalty $\rho$, the method returns a variety of solutions including the early and late fusion approaches for $\rho = 0$ and $\rho = 1$ respectively.

Cooperative learning has been proven to provide effective results in various scenarios, both including simulated and real multiomics data, specifically in the context of labor onset prediction \citep{Ding2022}.

\subsection{A multi-source model with interactions}

We consider two data views $X_1 \in \mathbb{R}^{n\times p_1}$, $X_2 \in \mathbb{R}^{n\times p_2}$, a common set of modifying variables $Z \in \mathbb{R}^{n \times K}$ and two sets of coefficients for main effect and interaction terms $\mathbf{\beta}^1 \in \mathbb{R}^{p_1}, \mathbf{\theta}^1 \in \mathbb{R}^{p_1 \times K}$ and $\mathbf{\beta}^2 \in \mathbb{R}^{p_2}, \mathbf{\theta}^2 \in \mathbb{R}^{p_2 \times K}$.
The model formulas for the pliable lasso in this context are:

$$ f_1(X_1,Z) = \sum_{j=1}^{p_1} X^1_j (\beta^1_j\mathbf{1} + Z\theta^1_j),$$
$$ f_2(X_2,Z) = \sum_{j=1}^{p_2} X^2_j (\beta^2_j\mathbf{1} + Z\theta^2_j).$$

Notice that the intercept has been omitted in our studies. In particular, we assume that the response $y$ has mean $0$: this has been implemented in all of the analysis that follow by centering $y$ by subtracting its mean. The modeling of the intercept in the general case could be implemented by introducing a single pair of coefficients $\beta_0$, $\theta_0$, common to the two data sources: this coefficients would be determined by regressing the residual on the matrix $(\mathbf{1}, Z)$. The optimization procedure when considering the intercept would be the same, just considering $y - \hat{\beta0} - Z\hat{\theta_0}$ instead of the centered response $y$.

In the context of cooperative learning, fixing the parameter $\rho$ we are looking to minimize the quantity:

\begin{equation} \label{eq1}
\begin{split}
L(\beta^1, \theta^1, \beta^2, \theta^2) & = \frac{1}{2} || \mathbf{y} - \sum_{j=1}^p X^1_j (\beta^1_j\mathbf{1} + Z\theta^1_j) - \sum_{j=1}^p X^2_j (\beta^2_j\mathbf{1} + Z\theta^2_j)||^2_2 \\
 & + \frac{\rho}{2}||\sum_{j=1}^p X^1_j \beta^1_j - \sum_{j=1}^p X^2_j \beta^2_j||^2_2\\
 & + (1-\alpha)\lambda\sum_{j=1}^{p_1}(||(\beta^1_j,\theta^1_j)||_2 + ||\theta^1_j||_2) + \alpha\lambda\sum_{j,k}|\theta^1_{jk}|\\
 & + (1-\alpha)\lambda\sum_{j=1}^{p_2}(||(\beta^2_j,\theta^2_j)||_2 + ||\theta^2_j||_2) + \alpha\lambda\sum_{j,k}|\theta^2_{jk}|.
\end{split}
\end{equation} 
The objective function can be broken down in three components:
\begin{enumerate}
\item
The first term represents the usual mean squared error, with $\mathbf{y} \in \mathbb{R}^n$ the target response vector, centered around $0$, from which the model's prediction is subtracted.
\item
The second term is the penalty that targets the alignment of the two models: note that we're not considering the difference of the two predictions in their entirety, but only considering the main effects and excluding the interaction terms. We make the assumption that both data sources have common modifying variables $Z$: when looking for agreement of the models, we then decide to leave the interaction effects unpenalized, focusing on alignment at the main effect level.
\item
The third and fourth term include the pliable lasso penalties for both models: note that we are considering a single value of $\lambda$ and $\alpha$ common to both views.
\end{enumerate}

As in \cite{Ding2022}, we compute the solution to the minimization problem (\ref{eq1}) through the following matrix adaptation. Letting:

\begin{equation}
\begin{split} \label{matrix_adap}
\tilde{X}=\left(\begin{array}{cc}X_1 & X_2 \\ -\sqrt{\rho} X_1 & \sqrt{\rho} X_2\end{array}\right), \tilde{Z}=\left(\begin{array}{l}Z \\ \mathbf{0}\end{array}\right)\\
\tilde{y}=\left(\begin{array}{l}y \\ 0\end{array}\right), \tilde{\beta}=\left(\begin{array}{c}\beta_1 \\ \beta_2\end{array}\right),  \tilde{\theta}=\left(\begin{array}{c}\theta_1 \\ \theta_2\end{array}\right),
\end{split}
\end{equation}
\vspace{0.3cm}

the equivalent problem becomes
$$L(\tilde{\beta}, \tilde{\theta}) = \frac{1}{2}||\tilde{y}- \tilde{X}\tilde{\beta}||_2^2 + (1-\alpha)\lambda\sum_{j=1}^{p_1+p_2}(||(\tilde{\beta_j},\tilde{\theta_j})||_2 + ||\tilde{\theta_j}||_2) + \alpha\lambda\sum_{j,k}|\tilde{\theta}_{jk}|.$$
which is in the same form as the pliable lasso, and can be solved as such through the R package \texttt{pliable} \citep{pliablemanual}. The package itself already contains the functions necessary to perform cross-validation to select the optimal $\lambda$ for each $\rho$. The process is repeated, keeping the same cross-validation folds, for each value of $\rho$: the parameter $\rho$ is then finally selected based on the minimum CV error out of the ones of optimal $\lambda$.
Algorithm \ref{alg1} outlines the procedure to fit the cooperative pliable model with the matrix adaptation method.\\

\begin{algorithm}
\caption{\textit{Algorithm for cooperative learning with pliable lasso.}}\label{alg1}
\begin{algorithmic}
\\
\textbf{Data:} $X_1 \in \mathbb{R}^{n\times p_1}, X_2 \in \mathbb{R}^{n\times p_2}, Z \in \mathbb{R}^{n\times K}$, the response $y \in \mathbb{R}^{n}$.
\vspace{0.1cm}
\For{$\rho \rightarrow  \rho_{\min}, \dots , \rho_{\max}$}
 \vspace{0.1cm}
  \begin{enumerate}
\item Set $\tilde{X}$, $\tilde{Z}$, $\tilde{y}$ as in Equation (\ref{matrix_adap}).
\item Solve the pliable lasso for $\tilde{X}, \tilde{Z}, \tilde{y} $, selecting the optimal $\lambda$ value through cross-validation.
    \end{enumerate}
\EndFor
\vspace{0.1cm}
\State Select the optimal value of $\rho$ based on the cross-validation error.
\end{algorithmic}
\end{algorithm}

The method presented so far only takes into account a single shrinkage parameter $\lambda$ for both data sources $X_1$ and $X_2$. In the case of sources that would require different amounts of shrinkage, the predictive ability can be improved by the inclusion of multiple hyper-parameters $\lambda_1$ and $\lambda_2$. This option, already considered for the original cooperative learning in \cite{Ding2022}, can be easily implemented by including penalty factors. Our version of the adaptive algorithm is reported in Algorithm \ref{alg_adap}.\\

\begin{algorithm}
\caption{\textit{Adaptive algorithm for cooperative learning with pliable lasso.}}\label{alg_adap}
\begin{algorithmic}
\\
\textbf{Data:} $X_1 \in \mathbb{R}^{n\times p_1}$, $X_2 \in \mathbb{R}^{n\times p_2}$, $Z \in \mathbb{R}^{n\times k}$, the response $y \in \mathbb{R}^{n}$.
\vspace{0.1cm}
\For{$\rho \rightarrow \rho_{min}, \dots , \rho_{max}$}
\vspace{0.1cm}
 \begin{enumerate}
     \item Solve the pliable lasso for $X_1, Z, y $, selecting the optimal $\lambda_1$ value through cross-validation. Repeat for $X_2, Z, y $, obtaining $\lambda_2$.
     \item Set $\tilde{X}$, $\tilde{Z}$, $\tilde{y}$ as in Equation (\ref{matrix_adap}).
\item Solve the pliable lasso for $\tilde{X}, \tilde{Z}, \tilde{y} $ with penalty factor $1$ for the first $p_1$ features, and $\frac{\lambda_2}{\lambda_1}$ for the other $p_2$, selecting the optimal $\lambda$ value through cross-validation.
 \end{enumerate}
\EndFor
\vspace{0.1cm}
\State Select the optimal value of $\rho$ based on the cross-validation error.
\end{algorithmic}
\end{algorithm}

\section{Simulation studies}\label{simulation_studies_section}

We want the simulated data to depend both on interaction terms and on the correlation between the data views: in order to achieve we adapt the latent factor model used in \cite{Ding2022} and include the role of simulated modifying variables. It is assumed that the features of both data sources and the modifying variables all share some underlying structure; in the example of omics data, this form of correlation would be the consequence of all data coming from the same patient.

In order to generate the data, we introduce: $n$ the number of data points, $p_1$, $p_2$ the number of features in the two data sources, $K$ the number of modifying variables, $p_u$ the number of latent factors (with $p_u < p_1$ and $p_u < p_2$), $t_1$ and $t_2$ the coefficients that regulate the prevalence of the latent factors in the composition of the two data views, and thus the correlation between the sources, $\sigma$, the variance of the error term, regulating the amount of noise present in the data. Furthermore, given the value for real coefficients $\beta^1$, $\theta^1$, $\beta^2$, $\theta^2$, we consider the following setup: 

\begin{enumerate}
\item Generate $x^1_j$, $x^2_j$, $z_k \in \mathbb{R}^n$ i.i.d as $\mathcal{N}(0,1)$.
\item For $i=1,2, \ldots, p_u$, generate $u_i \in \mathbb{R}^n$ i.i.d. as $\mathcal{N}(0,1)$ and define
\begin{enumerate}[label=(\roman*)]
    \item $x^1_i:=x^1_i+t_1\cdot u_i$
    \item $x^2_i:=x^1_i+t_2 \cdot u_i$
    \item if $i<K$, $z_i=z_i + u_i$
\end{enumerate}

\item Define $X^1=\left(x^1_1, x^1_2, \ldots, x^1_{p_1}\right)$,  
$X^2=\left(x^2_1, x^2_2, \ldots, x_{p_2}\right)$,  $Z=\left(z_1, z_2, \ldots, z_{K}\right)$.
\item Compute
$$\boldsymbol{y} = X^1\beta^1 + \sum_{j=1}^{p_1}(X^1_j \circ Z)\theta^1_j + X^2\beta^2 + \sum_{j=1}^{p_2}(X^2_j \circ Z)\theta^2_j + \epsilon,$$ where $\epsilon \in \mathbb{R}^n$ distributed i.i.d. as $N(0,\sigma^2).$
\end{enumerate}

\begin{figure}
\vspace{-1.5cm}
    \includegraphics[width=1.2\linewidth, center]{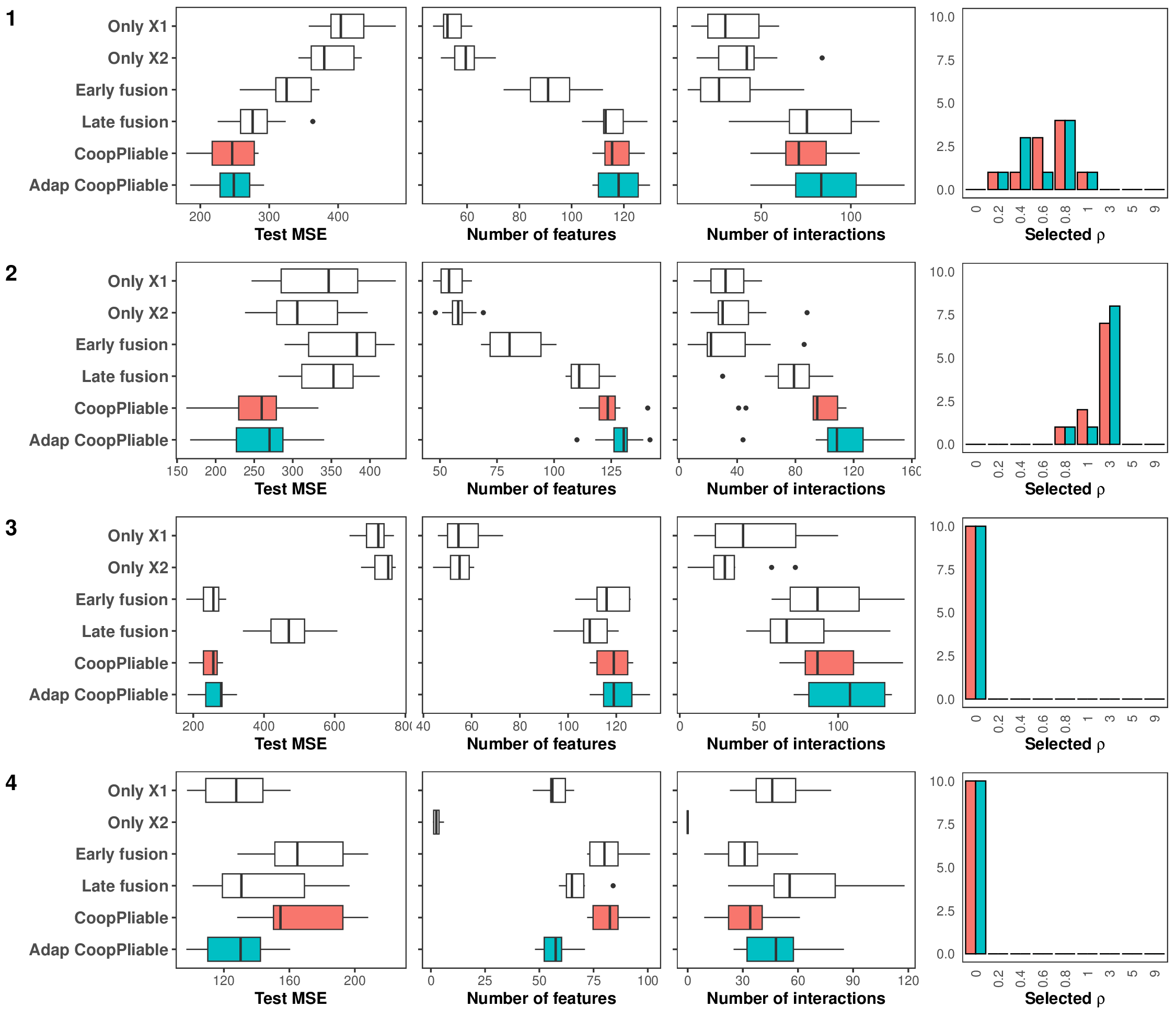}
    \caption[Simulation studies in the low dimensional setting. Test MSE, number of main effects and interactions and distribution of selected $\rho$.]{ \textit{Simulation studies in the low dimensional setting.} For each row, the first panel shows the comparison in test MSE between the methods, the second and third respectively the number of main effects and interaction terms selected in the best models, and the last one the distribution of selected $\rho$. Each experiment is repeated $10$ times. From top to bottom: (1) $X_1$ and $X_2$ have a medium level of correlation ($t_1 = t_2 = 2$), $SNR = 5.0$. (2) $X_1$ and $X_2$ have a high level of correlation ($t_1 = t_2 = 4$), $SNR = 1.7$. (3) $X_1$ and $X_2$ have no correlation, $SNR = 2.2$. (4) $X_1$ and $X_2$ have no correlation and only $X_1$ contains signal; in this case, the response is generated only including the component relative to data source $X_1$ ($t_1 = 2, t_2 = 0$), $SNR = 3.1$. The average MSE for the Only $X_2$ in this case is $2123.6$: its distribution is excluded from the plot for graphical purposes.}
    \label{simulation_figure1}
\end{figure}

\begin{figure}
\vspace{-1.5cm}
    \includegraphics[width=1.2\linewidth, center]{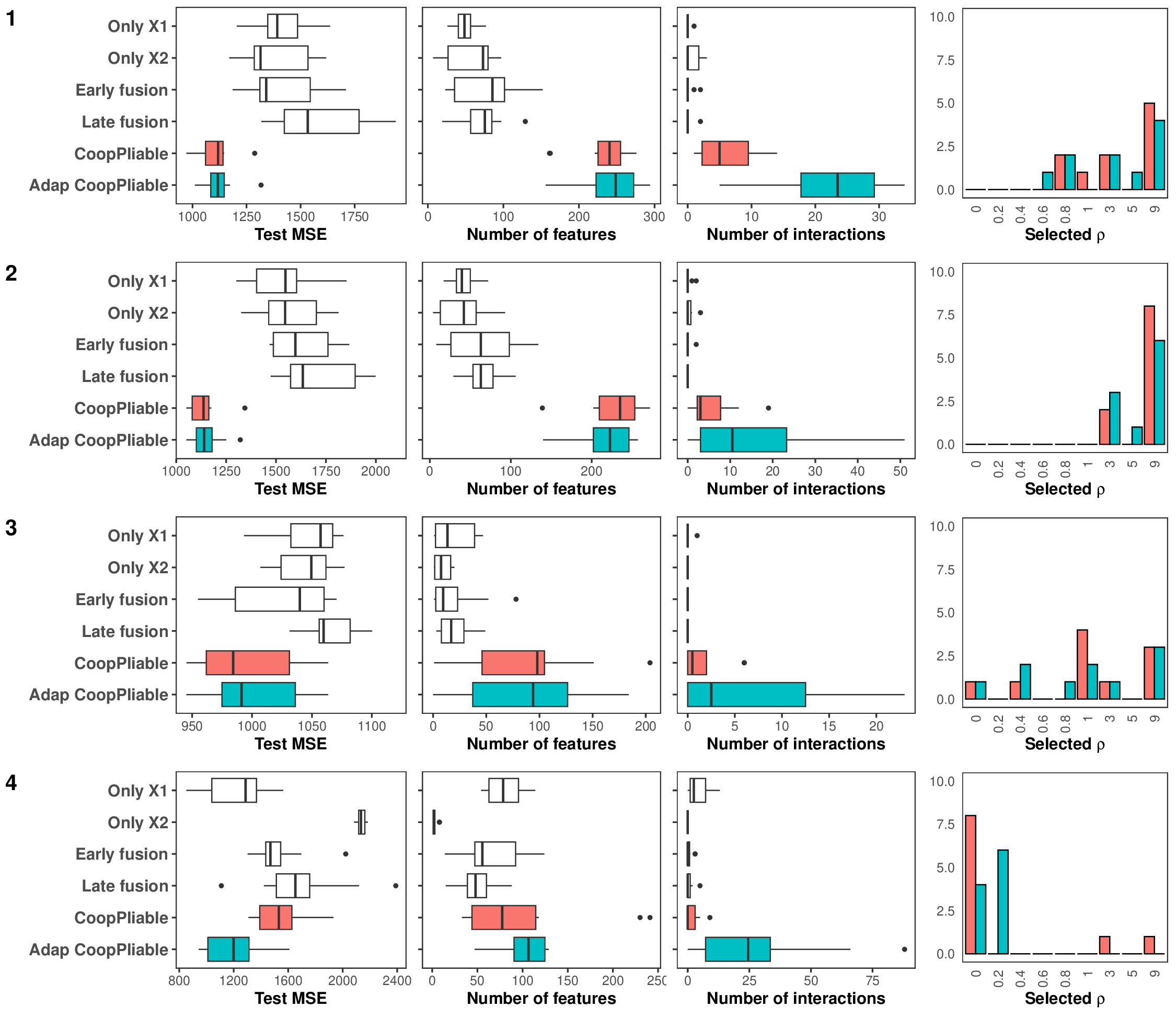}
    \caption[Simulation studies in the high dimensional setting. Test MSE, number of main effects and interactions and distribution of selected $\rho$]{ \textit{Simulation studies in the high dimensional setting.} For each row, the first panel shows the comparison in test MSE between the methods, the second and third respectively the number of main effects and interaction terms selected in the best models, and the last one the distribution of selected $\rho$. Each experiment is repeated $10$ times. From top to bottom: (1) $X_1$ and $X_2$ have a medium level of correlation ($t_1 = t_2 = 2$), $SNR = 2.4$. (2) $X_1$ and $X_2$ have a high level of correlation ($t_1 = 6, t_2 = 1$), $SNR = 1.6$. (3) $X_1$ and $X_2$ have no correlation, $SNR = 2.1$. (4) $X_1$ and $X_2$ have no correlation and only $X_1$ contains signal; in this case, the response is generated only including the component relative to data source $X_1$, $SNR = 3.5$.}
    \label{simulation_figure2}
\end{figure}

The comparison is made with the single source models (referred to as ``Only X1'' and ``Only X2''), early and late fusion and the cooperative method, both without and with the adaptive variation (``CoopPliable'' and ``Adap CoopPliable'' respectively).
In particular, early fusion is implemented by concatenating  $X_1$ and $X_2$ to obtain a combined matrix $X$, which is then used as the input for the pliable lasso. Late fusion results are obtained by fitting individual pliable lasso models from each distinct data source and combining the two separate predictions through a linear combination.
The results presented include Test MSE, as well as the the sparsity in the solution coefficients, both for main effects and interaction terms. Finally, we also show the distribution of the selected $\rho$, and highlight how this distribution changes with the parameters that describe the simulated data. Figure \ref{simulation_figure1} considers a low-dimensional setting. The training set contains $500$ datapoints and we set the number of features as $p_1=p_2=100$. Figure \ref{simulation_figure2} deals with the high-dimensional setting. The training set contains $200$ datapoints and we set the number of features as $p_1=p_2=500$. The size of the test set is $9800$. The studies explore different levels of correlation between the data sources, as well as different levels of noise.

The results of the simulations are summarized as follows:

\begin{itemize}
    \item \textbf{Test MSE}: when correlation is present between the two data sources (like in the first two simulations in Figure \ref{simulation_figure1} and \ref{simulation_figure2}), the cooperative methods are able to achieve lower test MSE results than the others considered, by selecting an optimal $\rho$ hyperparameter. When correlation is absent (like in the third simulation for each Figure), the methods can't leverage any shared information and fall back to a form of early fusion: they are still able to outperform the use of single sources when both of them contain signal. When only one source contains signal and the sources are not correlated, non-adaptive cooperative learning is outperformed by the separate model fit on the source containing the signal, and still performs on par with early fusion, while the adaptive method is able to perform on par with the separate model, outperforming early and late fusion.
    
    \item \textbf{Sparsity}: the cooperative methods lead to a less sparse model than the others considered, both in terms of main effects and interaction terms, especially when selecting higher values of the agreement penalty $\rho$. This is due to the presence of the agreement penalty, which reduces the impact of the pliable lasso penalty and its variable selection effect. We also notice that in the high-dimensional setting, where the magnitude of shrinkage introduced by the methods is generally higher, the separate fit models and early and late fusion hardly introduce any interaction terms at all.
    
    \item \textbf{Rho distribution}: the $\rho$ values considered in the experiments go from $0$ to $9$, with higher values of rho being selected in the presence of stronger correlation between the sources. The high-dimensional case also usually selects higher values of $\rho$. In the specific case of no correlation in the low-dimensional setting, $\rho$ was selected to be $0$ in every iteration: the cooperative method obtains the same results as a form of early fusion.
\end{itemize}

\subsection{Feature selection}
Other than evaluating performance by taking into consideration the test MSE, we want to compare the ability of the models considered to identify which variables are relevant to predicting the response and which are not. In particular we look at \textit{sensitivity} and \textit{specificity} for variable selection, which are computed as:\\

$$
\textrm{Sensitivity} = \frac{\textrm{True positives}}{\textrm{True positives} + \textrm{false negatives}},
$$

$$
\textrm{Specificity} = \frac{\textrm{True negatives}}{\textrm{True negatives} + \textrm{false positives}}.$$\\

\begin{figure}
\vspace{-1.5cm}
    \includegraphics[width=1\linewidth, center]{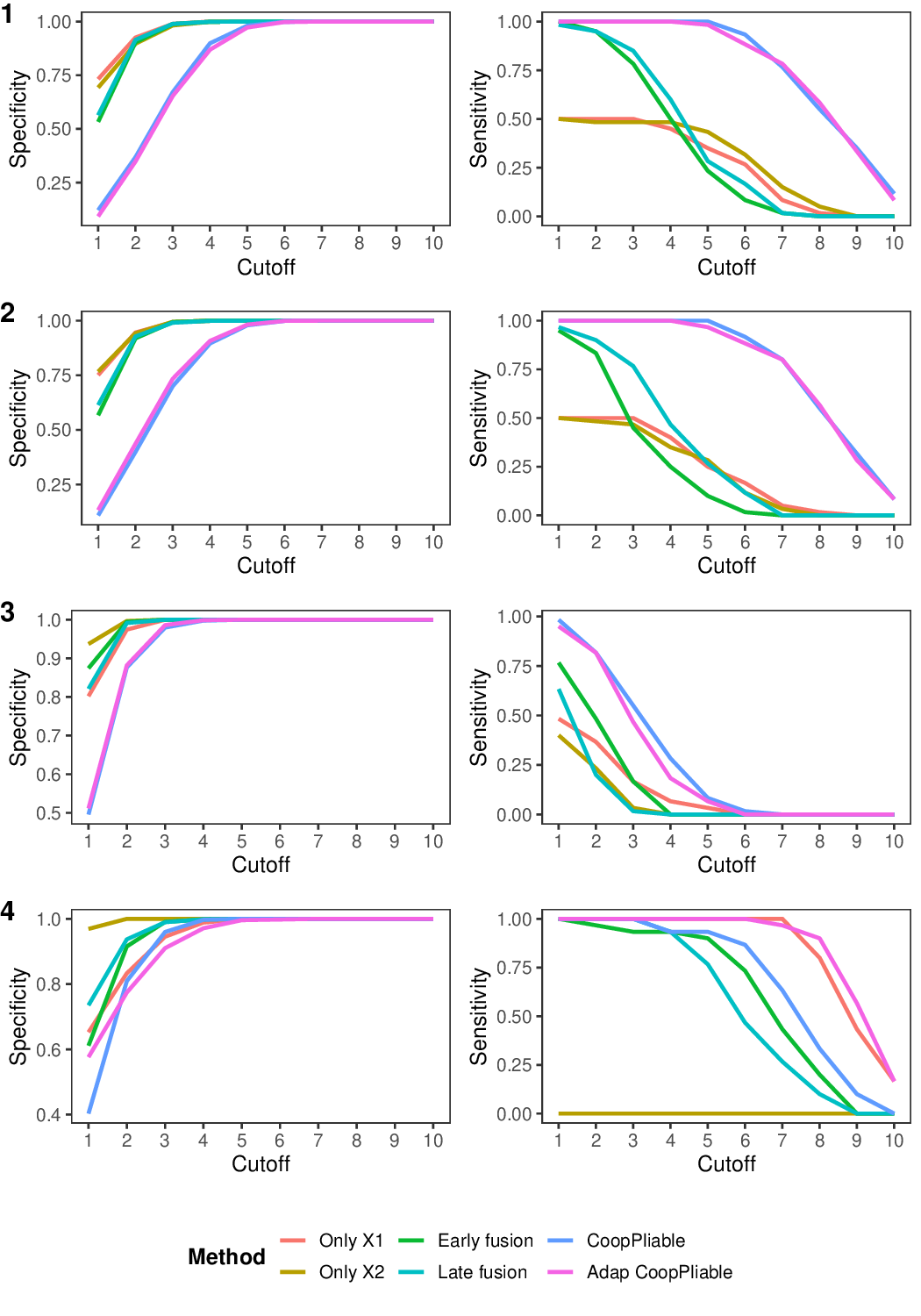}
    \caption{ \textit{Sensitivity and specificity results for different scenarios in high-dimensional simulations.} The settings are the same as in Figure \ref{simulation_figure2}: from top to bottom (1) $X_1$ and $X_2$ have a medium level of correlation ($t_1 = t_2 = 2$), $SNR = 2.4$. (2) $X_1$ and $X_2$ have a high level of correlation ($t_1 = 6, t_2 = 1$), $SNR = 1.6$. (3) $X_1$ and $X_2$ have no correlation, $SNR = 2.1$. (4) $X_1$ and $X_2$ have no correlation and only $X_1$ contains signal.}
    \label{sens_spec_plot}
\end{figure}

In particular, we consider the same simulation studies of Figure \ref{simulation_figure2}, repeat each simulation 10 times, and only consider selected the variables that are included in the models a number of times over a cutoff. 
Figure \ref{sens_spec_plot} presents the results obtained as the value of the cutoff changes. As we can see, the non-cooperative methods, tend to have higher specificity (correctly identifying which variables to exclude), up to a threshold of around $5$, where all methods converge to specificity $1$. In terms of sensitivity, the cooperative methods are consistently better: when looking at the rows where the sources share information (1 and 2), sensitivity remains almost perfect up until cutoff $6$. This advantage is not mantained when the sources are independent (row 3). We also note that, in the case of only one source containing signal (row 4), the use of Only X1 and the adaptive coop pliable method is able to outperform the rest in sensitivity.

\section{Real multi-omics studies}\label{section_real_multi-omics_studies}

In this section, the developed method has been applied to real datasets from two multi-omics settings: we consider a dataset for labor onset and one for cancer treatment response prediction.

\subsection{Labor onset data} \label{subsection_labor_onset}
We firstly applied the method to a data set of labor onset, already studied with the original cooperative learning method in \cite{Ding2022}: the data, as described in \cite{Stelzer2021}, was collected from a cohort of 63 women who went into spontaneous labor.
Current estimates for the onset of labor are based on the norm of a 40-week gestational period, but most pregnancies deviate from this duration. Identifying helpful biomarkers for determining the timing of delivery can inform a more comprehensive understanding of maternal biology in labor, and lead to more accurate predictions \citep{Stelzer2021}.

\begin{figure}
    \includegraphics[width=1.1\linewidth]{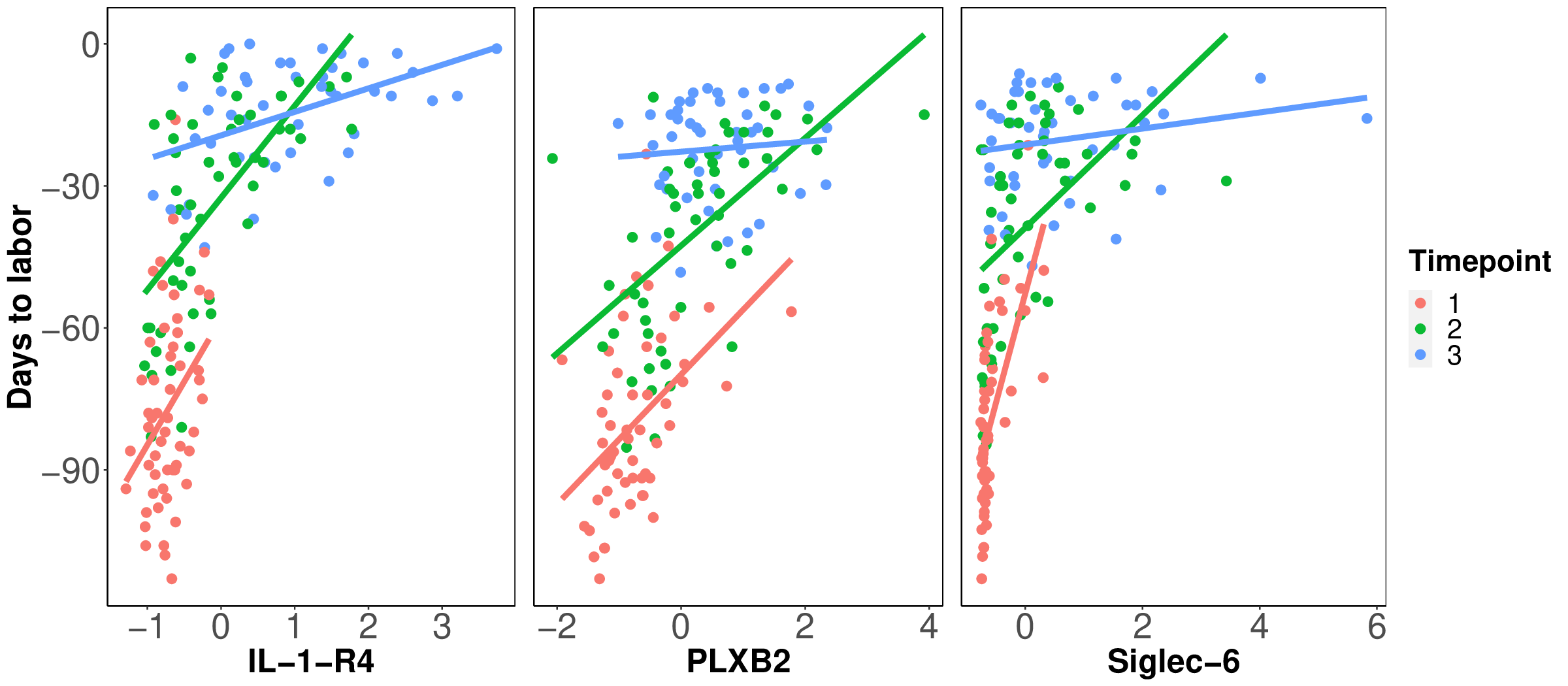}
    \caption{\textit{Scatterplots of three of the proteins in the dataset and the days to labor.} Linear regression lines are added to show the relationship between the variables and the response at each of the timepoints. For all variables, the relationship seems to be most strongly positive at timepoint 1 (in pink), and reduced for the following timepoints.}
    \label{labor_variables}
\end{figure}

As done in \cite{Ding2022}, the performed analysis aims to predict time to spontaneous labor onset using omics data. The data was obtained from blood samples collected from the patients during the last 120 days of pregnancy at three consecutive timepoints in a longitudinal study. Proteome and metabolome were used as the two main sources of data. 
The proteomics data contained measurements for 1,317 proteins and the metabolomics data contained measurements for 3,529 metabolites. We considered the numbered timepoint of blood sample collection as a modifying variable, and treated it as a categorical variable with categories $1$, $2$ and $3$ for the three time points for each patient.
The assumption behind this model is that the effects of relevant omic features for the prediction of labor onset can change through the evolution of pregnancy: the existing relations between proteome and metabolome predictors to the likelihood of labor may vary as parturition approaches. This is visualized in Figure \ref{labor_variables} for three of the proteins that were already identified as relevant in the prediction of labor onset from previous analyses of the same dataset: IL-1-R4, Plexin-B2 and sialic acid binding immunoglobulin-like lectin–6 (Siglec-6). As the scatterplots highlights, the variability of these biomarkers is especially large in the final period of pregnancy, thus the linear relationship between the variable and the response is not constant over the three timepoints, motivating the interest in introducing an interaction term with the timepoint variables in the model.

The use of multiple samples from each patient could present an issue, since it would introduce correlation between different data points. This problem was mitigated by splitting training and testing data based on patient, rather than on singular samples: if a patient was selected as part of the training set, their measurements at all timepoints would be in the training data and likewise for the test data.
The data was split in $80\%$ training and $20\%$ test samples. The analysis was conducted across $10$ different random splits of the data into training and test sets.\\

The results are presented in Table \ref{table_mse_timepoint}. The separate model considering only proteomics data achieves lower test MSE than the one fitted on metabolomics data, signifying that the proteomics data alone might have more predictive value than the metabolomics data alone.
The integration of both data sources is beneficial for early fusion, while late fusion performs worse than the model using only proteomics. The cooperative methods are able to outperform all other methods, giving test MSE gains over the model fit only on proteomics, as well as early and late fusion.
The number of main effects and interactions reported in Table \ref{table_mse_timepoint} is the rounded mean out of the $10$ iterations. When looking at the features included in the model, proteomics feature IL-1R4 (IL-1 receptor type 4) had the highest average coefficient for all methods (except when considering only the metabolome). This protein is described as an inhibitory receptor of the proinflammatory cytokine IL-33: its surge in the last phase of pregnancy could signify that IL-1R4 may be an important regulator of inflammation in this period and thus can be used as a predictor for labor onset \citep{Stelzer2021}.
Another highly rated feature from the proteome data is PLXB2 (Plexin-B2), which is a protein expressed by the fetal membranes \citep{Singh2014}. Both had been previously identified as important predictors. Other identified variables are Secretory Leukocyte Peptidase Inhibitor (SLPI) and Cystatin-C, which have both been found to have a potential role in placenta and cervix remodeling during pregnancy \cite{Lee2018, Samejima2021}.

\begin{table}[!h]
\begin{tabular}{lccc}
\toprule \textbf{Method} & \textbf{Test MSE (SD)} & \begin{tabular}{c} 
\textbf{Number of} \\
\textbf{main effects}
\end{tabular} & \begin{tabular}{c} 
\textbf{Number of} \\
\textbf{interactions}
\end{tabular} \\
\hline Only Proteomics & 544.65(66.75) & 56 & 6 \\
Only Metabolomics & 575.36(47.25) & 50 & 1 \\
Early fusion & 501.56(25.98) & 66 & 0 \\
Late fusion & 548.66(37.35) & 104 & 2 \\
\textbf{CoopPliable} & 439.81(28.34) & 127 & 4 \\
\textbf{AdapCoopPliable} & 410.32(18.01) & 124 & 54 \\
\bottomrule
\end{tabular}
\caption[Results on the labor onset data with timepoint as modifying variable. Test MSE, number of main effects and interactions.]{Results on the labor onset data with Timepoint as modifying variable.}
\label{table_mse_timepoint}
\end{table}

\subsection{Genomics of drug sensitivity in cancer data}

The Genomics of Drug Sensitivity in Cancer (GDSC) dataset is an extensive and highly utilized resource in the field of pharmacogenomics, as it comprises a vast collection of molecular and pharmacological data obtained from diverse cancer cell lines. Cell lines are laboratory-grown cells that originate from cancerous tumors and retain their characteristics, facilitating the study of cancer biology and treatment response. \citep{Mirabelli2019}. This dataset includes comprehensive genomic profiles of these cell lines, encompassing details on genetic mutations, gene expression levels, copy number alterations, and DNA methylation patterns as well as information on how these cell lines respond to a wide array of anticancer drugs as measured with cell viability assays. Further details on the data can be found in \cite{Yang2012}.\\

Our analysis focused on the sensitivity of 499 cell lines to the selected drug Nilotinib. This drug is a tyrosine kinase inhibitor that targets a protein encoded by the fusion gene called BCR-ABL, which is a known driver for initiation and maintenance of chronic myeloid leukemia (CML). By inhibiting the activity of this abnormal protein, it effectively suppresses the growth and proliferation of cancer cells in patients with CML. Nilotinib represents a significant advancement in cancer treatment, offering a targeted therapy that specifically acts on the molecular abnormality present in this particular type of cancer \citep{Sacha2019}.
In our analysis, the cell lines are characterized by two sources of genomic data. Firstly, 2602 gene expression features are pre-selected by selecting those features with the largest variances across cell lines, which in total explain 50\% of the variation. In addition, we use binary mutation data for 68 genes which are causally implicated in cancer according to the cancer gene census \citep{Zhao2020}.
Finally, the 499 cell lines represent tumour samples from 13 tissue types: we introduce this information binary dummy modifying variables that code the different types. The resulting interactions should highlight the different roles that genomic features play in predicting drug response based on the tissue type considered. Given the type of drug that we consider, which which targets chronic myeloid leukemia, we will expect most interactions to be selected with the variable representing blood as the tissue type. The analysis setup is similar to the one for the previous dataset, with 10 different random train-test splits of the data, with 80\% of datapoints for training and 20\% for testing.\\

\begin{figure}
\vspace{-3cm}
    \includegraphics[width=0.8\linewidth, center]{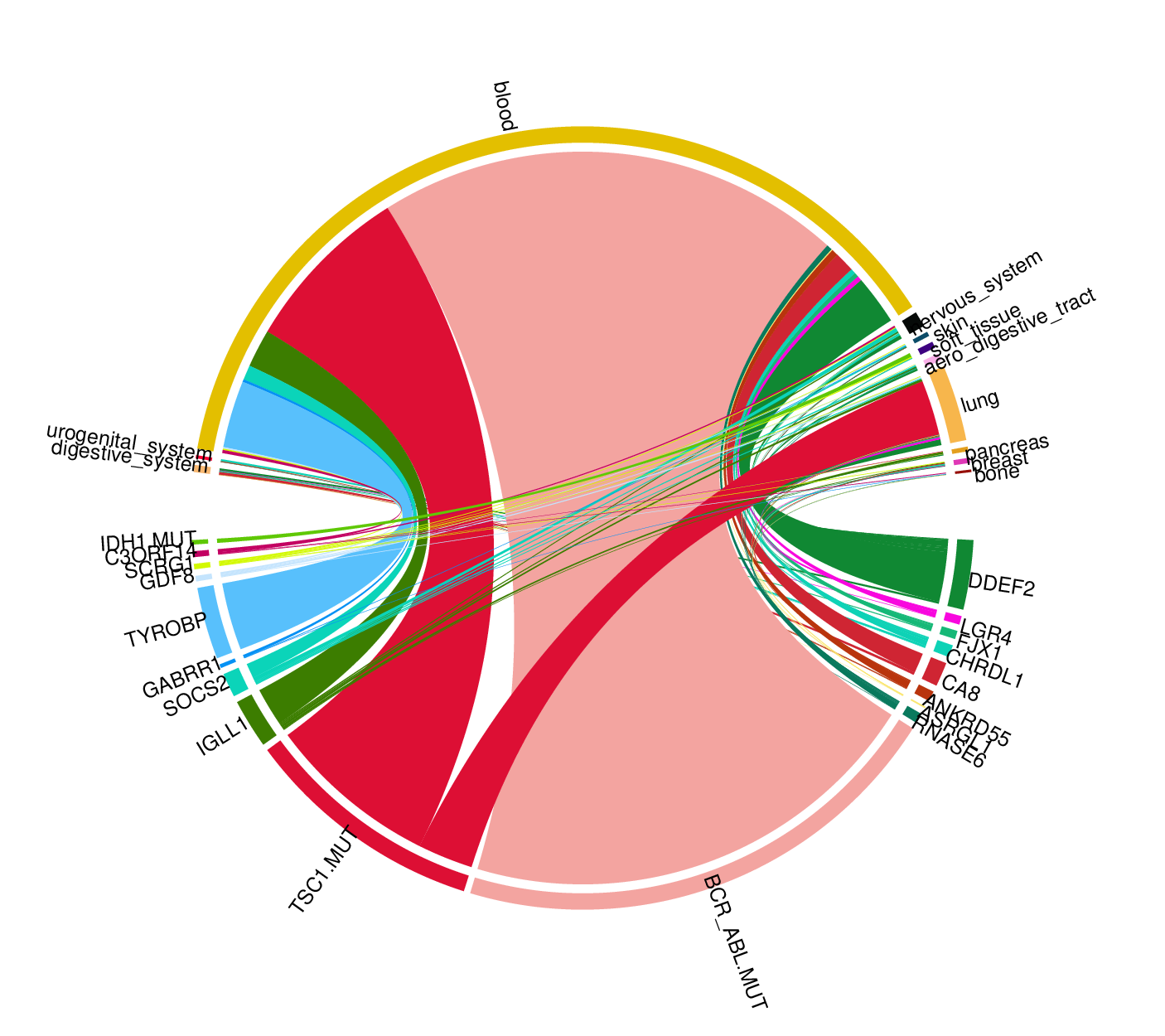}
    \caption{\textit{Chord diagram for the interaction effects between gene expression features and cancer types for Nilotinib.} The upper chord represents the cancer types and the lower chord represents the gene expression features. The width of the line connecting the cancer type to a gene indicates the strength of the interaction. For better readability, only the interactions that were very stable in the selection, being included in at least 8 out of 10 iterations, are plotted.}
    \label{chord_diagram_nilo}
    \vspace{1cm}
    \hspace{-0.25cm}
    \includegraphics[width=1\linewidth, center]{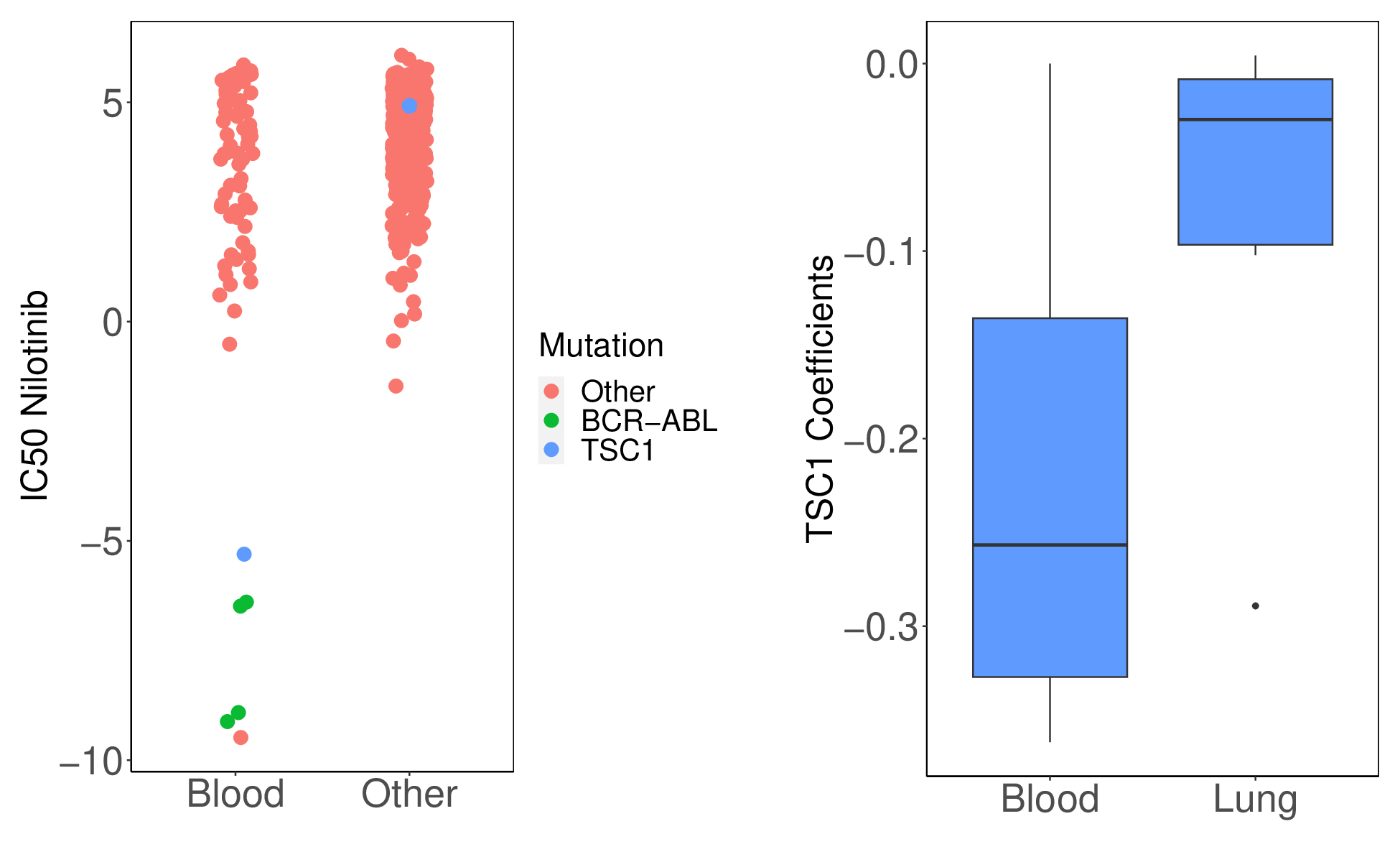}
    \caption{The mutation scatterplot (on the left) shows how the presence of BCR-ABL mutation is specific to tumours that react positively to Nilotinib, while the TSC1 mutation is present in a CML blood tumour case, as well as a lung tumour case that does not respond to Nilotinib. The boxplot represents the value of coefficient for the TSC1 mutation variable, when the interaction effects are accounted for.}
    \label{GDSC_interactions_plot}
\end{figure}

The results are presented in Table \ref{table_mse_GDSC_nilo}, with the cooperative methods returning the lowest test MSE. When considering the two sources separately, the model trained on only mutation data performs better than the one that only considers gene expression.
As we expected from the simulation results, cooperative learning includes more main effects and many more interactions when compared to the other methods, leading to a less sparse model: this is due to the presence of the agreement penalty, which reduces the variable selection effect of the pliable lasso penalty.
All methods (excluding the one trained only on gene expression) rank the presence of the BRC-ABL mutation as the most relevant variable, in terms of average absolute coefficient. This is consistent with our knowledge of the functioning of Nilotinib as discussed above, especially when considering the presence of interactions with the blood cancer type (see Figure \ref{chord_diagram_nilo}).
Another feature that is consistently included in the models is the TSC1 mutation, which is related to the tuberous sclerosis complex, a condition characterized by developmental problems and the growth of noncancerous (benign) tumors in many parts of the body \citep{Portocarrero2018}.

% \begin{table}[!h]
% \begin{tabular}{lccc}
% \hline \textbf{Method} & \textbf{Test MSE (SD)} & \begin{tabular}{c} 
% \textbf{Number of} \\
% \textbf{main effects}
% \end{tabular} & \begin{tabular}{c} 
% \textbf{Number of} \\
% \textbf{interactions}
% \end{tabular} \\
% \hline Only Proteomics & $544.65(66.75)$ & 40 & 0 \\
% Only Metabolomics & $575.36(47.25)$ & 29 & 0 \\
% Early fusion & $501.56(25.98)$ & 42 & 0 \\
% Late fusion & $548.66(37.35)$ & 42 & 8 \\
% \textbf{CoopPliable} & $439.81(28.34)$ & 94 & 1 \\
% \textbf{AdapCoopPliable} & $410.32(18.01)$ & 81 & 14 \\
% \hline
% \end{tabular}
% \caption[Results on the labor onset data with timepoint as modifying variable. Test MSE, number of main effects and interactions.]{Results on the labor onset data with Timepoint as modifying variable. The average $\rho$ selected is $2.52$.}
% \label{table_mse_timepoint}
% \end{table}

\begin{table}
\begin{tabular}{lccc}
\toprule \textbf{Method} & \textbf{Test MSE (SD)} & \begin{tabular}{c} 
\textbf{Number of} \\
\textbf{main effects}
\end{tabular} & \begin{tabular}{c} 
\textbf{Number of} \\
\textbf{interactions}
\end{tabular} \\
\hline
Only Gene Expression & 3.146 (0.297) & 65 & 19 \\
Only Mutations & 2.739 (0.288) & 9 & 3 \\
Early fusion & 2.712 (0.268) & 29 & 2 \\
Late fusion & 2.903 (0.291) & 10 & 3 \\
\textbf{CoopPliable} & 2.690 (0.268) & 128 & 221 \\
\textbf{AdapCoopPliable} & 2.663 (0.281) & 140 & 267 \\
\bottomrule
\end{tabular}
\caption[Results on the GDSC dataset considering the drug Nilotinib. Test MSE, number of main effects and interactions.]{Results on the GDSC dataset considering the drug Nilotinib.}
\label{table_mse_GDSC_nilo}
\end{table}

Further analyses on the interactions identified by the cooperative model are shown in Figures \ref{chord_diagram_nilo} and \ref{GDSC_interactions_plot}. The first plot shows the selected interactions between gene expression and mutation features and cancer types through a chord diagram.
In this example, we are showing the relationships between cancer types, represented by the upper chord, and gene expression and mutation features, represented by the lower chord. Other than the one with the BRC-ABL mutation, this plot also shows an interaction between the blood and lung cancer types and the TSC1 mutation. The second plot (Figure \ref{GDSC_interactions_plot}), gives a clearer outlook on the role of these interactions in our data example: the BCR-ABL mutation is present in 4 cell lines that present a much stronger response to Nilotinib than others, even in the blood cancer group. We note that in this case, since all BCR-ABL mutated cell lines belong to the same cancer type, the introduction of the interaction term is actually redundant, since it is not necessary in differentiating the coefficients between multiple cancer types. The TSC1 mutation, on the other hand, is present in two cell lines relative to blood and lung cancer respectively: the presence of interaction in this case can differentiate the predicted response of Nilotinib by modifying the related coefficient. In general, it is highlighted how these effects can help distinguish drug response in the specific case of CML tumours.

\section{Conclusion}\label{section_conclusion}

We present a statistical model for multi-source data that takes into account interactions. This is made possible by adapting the cooperative learning method \citep{Ding2022} by the introduction of a pliable lasso penalty \citep{Tibshirani2019}, that allows for the inclusion of interaction terms between the main features and a set of modifying variables. The adapted algorithm has been implemented in R and tested on both simulated and real data. The code used to perform all the simulations and the real data analysis is available at \url{https://github.com/matteodales/CooperativePliableLasso/}. The data-driven approach to multi-source integration allows for a better predictive performance than other data fusion methods, especially when there are correlations present between the data sources. The procedure generally leads to a less sparse model both in selection of main effects and interactions, leading to a reduction in specificity of selection but improving sensitivity. We consider two real multiomics application examples: the first deals with predicting time to labor onset from proteomics and metabolomics data, and the second aims at predicting treatment response for cancer cell lines based on gene expression and mutation data in a large in-vitro pharmacogenomic screen. In both cases our method was able to obtain lower prediction error, and we could show examples of relevant features and interactions being identified.

Several extensions of the proposed approach could be explored in the future. The use of the pliable lasso for the inclusion of interactions is advantageous when there is a pre-specified set of modifying variables that we assume might influence the relationship between the main effects and the response, but doesn't allow for the modeling of more general interactions between the variables in each source. Given the flexibility of cooperative learning in terms of fitting methods, this approach could be adapted to consider other examples of hierarchical interaction models such as in \cite{Lim2015} or \cite{Haris2016}. If we are also interested in interactions between features in different sources, \cite{Ding2022} suggests a variation of cooperative learning with an added penalty term that depends on all features combined.
Furthermore, the main focus of the method is on prediction, and no variable selection guarantees are yet in place. In order to be able to better make use of the method presented by providing statistical guarantees on the identified variables and interactions, some approaches of false discovery control could be implemented. This would allow for the reduction of the number of predictors erroneously deemed significant. For example, the use of a form of stability selection \citep{Meinshausen2010} could be explored.
Finally, the multiomics datasets we included are only illustrative examples aimed at assessing the performance of the method in real settings. However, its applicability is in principle very large, and the investigation of a larger variety of datasets, possibly with the inclusion of more than two data sources, would be beneficial in understanding its capabilities.

\bibliographystyle{unsrtnat}
%\bibliography{‎⁨‎⁨‎⁨‎⁨‎⁨references} % <========================================
\bibliography{sample}

\end{document}